\documentclass[twocolumn,aps,amsmath,amssymb,prb]{revtex4}
\usepackage{graphicx}

\renewcommand{\L}{{\cal L}}

\newcommand{\sK}{{\sf K}}

\newcommand{\sT}{{\sf T}}
\newcommand{\bfP}{\mbox{\boldmath $P$}}
\newcommand{\bfC}{\mbox{\boldmath $C$}}

\newcommand{\Sec}[1]{Sec.\,\ref{#1}}
\newcommand{\App}[1]{Appendix\,\ref{#1}}
\newcommand{\ti}{\tilde}

\newcommand{\be}{\begin{equation}}
\newcommand{\ee}{\end{equation}}
\newcommand{\bea}{\begin{eqnarray}}
\newcommand{\eea}{\end{eqnarray}}
\newcommand{\bsube}{\begin{subequations}}
\newcommand{\esube}{\end{subequations}}
\newcommand{\Eq}[1]{eq\,\ref{#1}}
\newcommand{\Eqs}[1]{eqs\,\ref{#1}}
\newcommand{\Fig}[1]{Fig.\,\ref{#1}}

\newcommand{\la}{\langle}
\newcommand{\ra}{\rangle}

\newcommand{\kT}{\mbox{$k_{\rm B}T$}}
\newcommand{\NA}{\mbox{\tiny NA}}

\begin{document}

\title{
  The quantum solvation, adiabatic versus nonadiabatic, and Markovian versus
  non-Markovian nature of electron transfer rate processes
}

\author{Rui-Xue Xu,$^{a)\ast}$
         Ying Chen,$^{a)}$
          Ping Cui,$^{a,b)}$
       Hong-Wei Ke,$^{b)}$
}
 \author{YiJing Yan$^{a,b)}$}
\email{rxxu@ustc.edu.cn; yyan@ust.hk}

 \affiliation{$^{a)}$Hefei National Laboratory for Physical Sciences
  at Microscale, University of Science and Technology of China,
  Hefei, Anhui, 230026, China    \\
  $^{b)}$Department of Chemistry,
    Hong Kong University of Science and Technology, Kowloon, Hong Kong}
\date{\today}

\begin{abstract}
  In this work, we revisit the electron transfer rate theory,
 with particular interests in the distinct quantum solvation effect,
 and the characterizations of adiabatic/nonadiabatic
 and Markovian/non-Markovian rate processes.
 We first present a full account for
 the quantum solvation effect on the electron
 transfer in Debye solvents,
 addressed previously in
 J.\  Theore.\  \&  Comput.\  Chem.\  {\bf 5}, 685 (2006).
  Distinct reaction mechanisms,
 including  the quantum solvation-induced transitions
 from barrier-crossing to tunneling, and
 from barrierless to quantum barrier-crossing rate processes,
 are shown in the fast modulation or low viscosity regime.
 This regime is also found in favor
 of nonadiabatic rate processes.
 We further propose to use Kubo's motional narrowing
 line shape function to describe the Markovian character of the reaction.
 It is found that a non-Markovian rate process
 is most likely to occur
 in a symmetric system in the fast modulation regime,
 where the electron transfer is dominant by tunneling due to
 the Fermi resonance.
\end{abstract}

\maketitle

\section{Introduction}
\label{thintro}

\subsection{Prelude}
\label{thintroA}
 Electron transfer (ET) is the simplest reaction system
but plays a pivotal role in many chemical and biological
processes. The field of ET research has grown enormously since
1950s.\cite{Mar56966,Mar64155,Mar85265,Kes742148,Jor80193,Zus80295,Zus8329,%
Gar854491,Fra85337,Wol871957,Spa873938,Spa883263,Spa884300,%
Yan884842,Yan896991,Muk89301,Yan979361,Tan966,Tan973485,Rip87411,Bix9935}
  The standard ET system--bath model Hamiltonian reads
 \be \label{HT}
 H_{\rm T} = h_a |a\ra\la a| + (h_b+E^{\circ})|b\ra\la b|
     + V(|a\ra\la b|+|b\ra\la a|).
 \ee
 Here, $E^{\circ}$ denotes the reaction endothermicity,
 $V$ the transfer coupling matrix element, and $h_a$ ($h_b$) the
 solvent Hamiltonian for the
 ET system in the donor (acceptor) state.
 The system is initially in the donor $|a\ra$ site, with
 the solvent (bath) equilibrium density matrix
 $\rho_a^{\rm eq}\propto e^{-h_a/(k_{\rm B}T)}$ at
 the specified temperature $T$.

   The ET system of \Eq{HT} can be treated as a
 spin-boson problem in the context
 of quantum dissipation. We have recently
 constructed a general theory of
 quantum dissipation,\cite{Xu05041103,Xu07031107,Jin07134113}
 which results in an analytical solution to
 the ET dynamics in a Debye (solvent)
 dissipation.\cite{Han0611438,Han06685}
 It is noticed that the exact construction leads always
 to a generalized rate
 equation:\cite{Nit06,Han0611438,Han06685}
 \be\label{copt}
  \dot P_a(t) =
 -\int_0^t\!\!d\tau  \hat  k(t-\tau) P_a(\tau)
 +\int_0^t\!\!d\tau  \hat k'(t-\tau) P_b(\tau).
 \ee
 Here, $\hat k(t)$ and $\hat k'(t)$ denote the forward
  and backward rate memory kernels.
 On the other hand, one often finds in practice that the simple
  kinetic theory, with a Markovian rate constant description,
 can well describe the observed rate process.
 The resulting reactant population $P_a(t)$ decays exponentially
 toward its equilibrium value at a given temperature.
  Note that formally the rate constant is just the
 integrated rate kernel over time.
 Is there any quantitative justification
 for the Markovian rate processes being
 often observed experimentally?
 In this work, we try to address this issue,
 along with the quantum solvation effect and
 the adiabatic versus nonadiabatic nature
 of ET rate processes.

\subsection{Background}
\label{thintroB}

  Previous work on ET theory focuses mainly on
 the rate constant description. Consider \Fig{fig1}, the schematics of
 the simple donor-acceptor ET system of \Eq{HT}. Here,
 the relevant (macroscopic) solvent potential surfaces
 $V_a$ and $V_b$ are plotted as the functions of solvation
 coordinate $U\equiv h_b-h_a$.
 The solvation energy for the ET from the
 donor $|a\ra$ to the acceptor $|b\ra$ site
 is given by \cite{Mar56966,Mar64155,Mar85265}
 \be \label{lamb}
  \lambda\equiv {\rm tr}_{\rm B} \left(U\rho_a^{\rm eq}\right)
 \equiv \la U \ra.
 \ee
 The second identity here defines the notation $\la\cdots\ra$
 for the bath ensemble average, where the trace runs over all
 the solvent (bath) degrees of freedom.
  At the crossing ($U+E^{\circ}=0$) point,
 $V_a=V_b=(E^{\circ}+\lambda)^2/(4\lambda)$.
 It is the celebrated Marcus'
 ET reaction barrier height.\cite{Mar56966,Mar64155,Mar85265}
 Thus, \Fig{fig1} also summarizes
 the Marcus' nonadiabatic rate expression,
 \be\label{MarcusRate}
  k_{\mbox{\tiny NA}}
    = \frac{V^2/\hbar}{\sqrt{\lambda k_{\rm B}T/\pi}}
   \exp\Big[-\frac{(E^{\circ}+\lambda)^2}{4\lambda k_{\rm B}T}\Big].
 \ee
 This rate (constant) can be derived readily from \Eq{HT}, by using the
 classical or static Franck-Condon approximation,
 followed by the classical fluctuation-dissipation theorem,
 $\la U^2 \ra - \la U \ra^2 = 2k_{\rm B}T\lambda$.
 It does not consider the {\it dynamic} solvation
 effect, which is characterized by the relaxation time and
 also associated with the viscosity of the
 solvent.\cite{Spa873938,Spa883263,%
  Spa884300,Yan884842,Muk89301,Yan896991}
   Such effect was first studied by Kramers in his
 classical Fokker-Planck-equation approach to the rate theory of
 isomerization reaction.\cite{Kra40284}
 The resulting rate shows
 the celebrated turnover behavior that
 the rate has a maximum in an intermediate viscosity
 region.\cite{Kra40284,Han90251}
 This clearly demonstrates the dual role of solvent on reaction rate.
 Note that in the Kramers' rate theory, the reaction
 system is treated as a Brownian particle moving on
 a single adiabatic double-well potential surface.

  The dynamic solvation on nonadiabatic ET
 has been incorporated in the quantum extension of Marcus' theory,
 formulated via the Fermi golden rule.\cite{Tan966,Tan973485,Rip87411,Bix9935}
 The solvation coordinator is now
 a time-dependent stochastic operator
 $U(t) \equiv e^{ih_at/\hbar}Ue^{-ih_at/\hbar}$,
 and assumed to follow the Gaussian statistics.
  Thus, the dynamic solvation is completely characterized
 by the correlation function,
 \be \label{Ct}
    C(t-\tau) = \la [U(t)-\lambda][U(\tau)-\lambda] \ra.
 \ee
 The Fermi-golden-rule formalism is valid for an
 arbitrary form of the solvation correlation function.
 But like the Marcus' theory, it is restricted
 in the second-order transfer coupling regime.
 The resulting rate does not show the Kramers' fall-off
 behavior\cite{Kra40284,Han90251} that
 involves the barrier recrossing and thus
 depends on higher-order transfer coupling.
  The improved approach has been proposed,
 on the basis of the fourth-order perturbation theory,
 followed by certain resummation schemes.\cite{Hyn85573,%
 Gar854491,Fra85337,Wol871957,Spa873938,Spa883263,%
 Spa884300,Yan884842,Muk89301,Yan896991,Yan979361}
  The resulting rate does support a smooth transition
 between nonadiabatic and adiabatic ET reaction,
 recovering properly the Kramers' turnover behavior.\cite{Kra40284,Han90251}

   In this work, we adopt the reduced density matrix
 dynamics approach that is closely related to our recent
 development of quantum dissipation
 theory.\cite{Xu05041103,Xu07031107,Jin07134113}
 With the aid of the analytical expression for the
 nonperturbative and non-Markovian ET rate,\cite{Han0611438,Han06685}
 we will elaborate in detail the
 quantum solvation effects in both the adiabatic and nonadiabatic
 reaction regimes, and further investigate
 their relations to the Markovian versus non-Markovian
 nature of ET reaction kinetics.

  The remainder of this paper is organized as follows.
 Section \ref{thsec2} reviews the generalized kinetic rate theory,
 constructed readily via the reduced density matrix dynamics.
  In \Sec{thsec3} we decompose the total
 rate into its adiabatic and nonadiabatic
 components, and also analyze the Markovian versus
 non-Markovian nature of population transfer dynamics.
 We propose to use the Kubo's motional narrowing function,
 originally used in the context of
 optical spectroscopy,\cite{Kub66255,Kub69101}
 to characterize the Markovian character of  ET rate process.
 Section \ref{thnum1} presents a full account of
 the effect of quantum versus classical solvation
 on some representative ET reaction systems.
 Section \ref{thnum2} analyses the population
 transfer dynamics, with the focus on their adiabatic/nonadiabatic
 and Markovian/non-Markovian characters.
 Finally, \Sec{thsum} concludes the paper.

\section{Non-Markovian rates: Theory}
\label{thsec2}

\subsection{Generalized rate theory and dissipation}
\label{thsec2A}

  Let us define the dynamic rates via the evolution
of reduced system density matrix $\rho$.
In the absence of time-dependent external field,
the generalized quantum master equation reads,
\be \label{copini}
 \dot\rho(t)=-i\L\rho(t) -
  \int_{0}^t\!\! d\tau \hat\Pi(t-\tau)\rho(\tau).
\ee
${\cal L}\rho\equiv \hbar^{-1}[H,\rho]$, with
$H\equiv\la H_{\rm T}\ra$ being the
 reduced system Hamiltonian;
$\hat\Pi(t-\tau)$ denotes the dissipation kernel.
Note that \Eq{copini} is formally exact; in fact,
the involving nonperturbative $\hat\Pi(t)$ can be
generally expressed in terms of continued
fraction formalism.\cite{Xu07031107,Tan89101,Tan06082001}

 To obtain the dynamic rates, we shall eliminate
the coherent components $\rho_{jk}$; $j\neq k$,
from \Eq{copini}, and retain the
populations $P_{j}(t)\equiv\rho_{jj}(t)$ only.
To do that, let us recast \Eq{copini} in its
Laplace frequency-domain resolution,
\be\label{rhos}
 s\ti\rho(s)-\rho(0)=-[i{\cal L}+\Pi(s)]\ti\rho(s),
\ee
and then arrange it into the block matrix form
for the population and coherence vectors,
$\ti\bfP=\{\ti\rho_{jj}\}$ and $\ti\bfC=\{\ti\rho_{jk};j\neq k\}$,
with $\bfC(t=0)={\bf 0}$, respectively. We have
\bsube\label{rhosmat}
\be \label{rhosmatP}
  s\ti\bfP(s) - \bfP(0)
 = -{\sT}_{\mbox{\tiny PP}}(s)\ti\bfP(s)
   -{\sT}_{\mbox{\tiny PC}}(s)\ti\bfC(s),
\ee
\be \label{rhosmatC}
  s\ti\bfC(s)
  = - {\sT}_{\mbox{\tiny CP}}(s)\ti\bfP(s)
  - {\sT}_{\mbox{\tiny CC}}(s)\ti\bfC(s),
\ee
\esube
with the transfer matrices,
${\sT}_{\mbox{\tiny PP}}$,
${\sT}_{\mbox{\tiny PC}}$, ${\sT}_{\mbox{\tiny CP}}$,
and ${\sT}_{\mbox{\tiny CC}}$, arising from
the corresponding rearrangement of $i\L+\Pi(s)$.
Eliminating the coherent component $\ti\bfC$ leads to
\bsube\label{PvecsNt}
\be \label{Pvecs}
  s\ti{\mbox{\boldmath $P$}}(s)-{\mbox{\boldmath $P$}}(0)
 =\sK(s)\ti{\mbox{\boldmath $P$}}(s),
\ee
or, equivalently,
\be \label{ratePt0}
 \dot P_j(t)=\sum_k \int_0^t\!d\tau \hat{\sK}_{jk}(t-\tau) P_k(\tau).
\ee
\esube
The involving rate resolution matrix is obtained as
\be \label{rateKs}
 \sK(s) \equiv \int_{0}^{\infty}\!\!dt e^{-st} \hat{\sK}(t)
   =    {\sT}_{\mbox{\tiny PC}}(s+{\sT}_{\mbox{\tiny CC}})^{-1}
    {\sT}_{\mbox{\tiny CP}} - {\sT}_{\mbox{\tiny PP}} .
\ee
Its element ${\sK}_{jk}(s)$ resolves the
state-to-state dynamic rate memory kernel $\hat\sK_{jk}(t)$.
The rate matrix satisfies
the relation $\sum_j \sK_{jk}=0$,
as inferred from the population
conservation, ${\rm Tr}\rho = \sum_j P_j = $ constant.

\subsection{Generalized rate theory in the two-state system}
\label{thsec2B}
 For the two-state ET system (cf.\ \Eq{HT} and \Fig{fig1})
of present study, only a single dynamic rate equation
is independent. That is \Eq{copt}, which reads in the Laplace
 domain as
\be\label{cop}
 s\ti P_a(s)-P_a(t=0)=-k(s)\ti P_a(s)+k'(s)\ti P_b(s).
\ee
 Here, $k(s)\equiv L\{\hat k(t)\}=-\sK_{aa}(s)$ and
 $k'(s)\equiv L\{\hat k'(t)\}=\sK_{ba}(s)$
 are the forward and backward rate resolutions, respectively.
 The involving transfer matrices for the
 simple ET system of \Eq{HT} have all been identified;
 cf.\ the eqs 32 of Ref.\ \onlinecite{Han0611438}:
\bsube \label{allT}
\be \label{allTa}
 \sT_{\mbox{\tiny PP}}=0,
\ \ \
  \sT_{\mbox{\tiny PC}} =
    \frac{iV}{\hbar}\left[\begin{array}{cc}
    -1 & 1 \\ 1 & -1
   \end{array}\right],
\ee
\be\label{allTb}
\ \
 \sT_{\mbox{\tiny CP}} =
     \left[\begin{array}{cc}
     -iV/\hbar & z^{\ast} +iV/\hbar \\ iV/\hbar & z-iV/\hbar
  \end{array}\right],
\ee
\be\label{allTc}
  \sT_{\mbox{\tiny CC}}
 = \left[\begin{array}{cc}
     x^{\ast} -i(E^{\circ}+\lambda)/\hbar & y^{\ast}
 \\
     y & x+i(E^{\circ}+\lambda)/\hbar
   \end{array} \right] .
\ee
\esube
Here,
\be\label{xyz}
 x \equiv \Pi_{ba,ba}, \ \
 y \equiv \Pi_{ba,ab}, \ \
 z \equiv \Pi_{ba,bb},
\ee
and their complex conjugates $\Pi^{\ast}_{jj',kk'}=\Pi_{j'j,k'k}$
are the only nonzero elements of the $\Pi$-tensor.
Denote also
\be\label{alpha}
 \alpha(s) \equiv  s+x(s)+(i/\hbar)(E^{\circ}+\lambda).
\ee
 The final expressions for the
rate resolutions are then\cite{Han0611438}
\bsube\label{finalK}
\be \label{Ks}
  k(s) = \frac{2|V|^2}{\hbar^2 }{\rm Re}\left\{
    \frac{\alpha(s) + y(s)}{|\alpha(s)|^2 - |y(s)|^2}\right\},
 \ee
  and
 \be \label{Kbs}
  k'(s) = \frac{2|V|^2}{\hbar^2}
    {\rm Re} \left\{
 \frac{[\alpha(s) + y(s)][1-i\hbar z^{\ast}(s)/V]}
   {|\alpha(s)|^2 - |y(s)|^2} \right\}.
 \ee
\esube
 The involving parameters, defined in \Eq{xyz}, can in principle
be evaluated in terms of continued fraction formalism
of the nonperturbative dynamics of
 reduced density matrix.\cite{Xu07031107,Tan89101,Tan06082001}
In particular, the continued fraction expressions
 of these parameters
have been solved analytically for the
Debye solvent model, where
the solvation correlation assumes a
 single exponential form.\cite{Han0611438}

\section{Nature of rate processes: Adiabatic versus nonadiabatic
 and Markovian versus non-Markovian}
\label{thsec3}

\subsection{Adiabatic--nonadiabatic rate decomposition}
\label{thsec3A}
  It is noticed that the ET in the short-time ($t<\hbar/V$) regime
 is always nonadiabatic, or $k\propto V^2$.
 This fact can be inferred from the observation that
 the backscattering events,
 responsible by the higher-order transfer coupling,
 are yet not to occur.
 In \App{thapp_perturate}, we solve
 \Eqs{rhosmat} and \ref{allT} in the weak transfer coupling limit,
 and obtain the nonadiabatic counterpart of \Eq{finalK}.
 \bsube\label{kNAall}
 \be\label{kNAs}
  k_{\NA}(s)=\frac{2V^2}{\hbar^2}
    {\rm Re}\!\int_0^{\infty}\!\!dt\, \exp[-st-g(t)],
\ee
with
\be \label{gfunc}
  g(t)=\frac{i}{\hbar}(E^{\circ}+\lambda)t+\frac{1}{\hbar^2}\!
   \int_0^t\!\!d\tau\!\int_0^{\tau}\!\!d\tau' C(\tau').
\ee
\esube
 The corresponding $k_{\NA}(s=0)$ is the
 well established quantum nonadiabatic
 rate expression.\cite{Hyn85573,%
 Gar854491,Fra85337,Wol871957,Spa873938,Spa883263,Spa884300,%
 Yan884842,Yan896991,Muk89301,Yan979361,Tan966,Tan973485,Rip87411,Bix9935}
 The Marcus' expression, \Eq{MarcusRate}, can be
 obtained via setting $C(t)\approx C(0)=2\lambda\kT$ that
 amounts to the static (slow-modulation)  approximation,
 followed by the classical fluctuation-dissipation theorem.

   Consider now the long-time regime, involving only the
 integrated rate kernel, i.e., the rate constant
 $k\equiv k(s=0)$ or $k'\equiv k'(s=0)$.
 Hereafter, rate constant will be referred in short as rate
 if it causes no confusion.
 For the later use, denote also the reaction equilibrium constant,
\be\label{Keq}
 K_{\rm eq} = \frac{P_{b}(\infty)}{P_a(\infty)}
  = \frac{k(s=0)}{k'(s=0)} \equiv \frac{k}{k'}.
\ee

  It is noticed that $k\leq k_{\NA}$ is expected to hold
 in general; see \Fig{fig6} and its comments followed, especially
 on the case of exception.
 This inequality may be inferred from
 the fact that the total $k$ contains the
 backscattering contributions, while
 the nonadiabatic $k_{\NA}$ does not.
 The decomposition of the total rate into
 its nonadiabatic and adiabatic components
 may therefore be introduced as\cite{Spa873938,Spa883263,Spa884300,%
  Yan884842,Yan896991,Muk89301}
\be \label{kA_def}
  \frac{1}{k} = \frac{1}{k_{\NA}} + \frac{1}{k_{\mbox{\tiny A}}}.
\ee
 As $k$ and $k_{\NA}$ can be evaluated via \Eq{Ks} and \Eq{kNAall},
 respectively, with $s=0$, the above equation
 can in fact be considered as the working definition
 of adiabatic $k_{\mbox{\tiny A}}$.
 We shall show later that $k_{\mbox{\tiny A}}$
 is  relatively much insensitive (compared with $k_{\NA}$)
 to the transfer coupling strength $V$ variable (cf.\ \Fig{fig5}).
 Therefore, it can be practically used to describe the
 adiabatic rate process that involves only the
 ground solvation surface via the diagonalization of \Eq{HT}.
 The ratio $k_{\NA}/k{\mbox{\tiny A}}$ can be considered
 as the adiabaticity parameter. The ET reaction assumes adiabatic
 when $k_{\NA}/k{\mbox{\tiny A}}\gg 1$,
 and nonadiabatic when $k_{\NA}/k{\mbox{\tiny A}}\ll 1$.

\subsection{Characterization of Markovian versus non-Markovian rate processes}
\label{thsec3B}
 We now turn to the Markovian versus non-Markovian nature
 of the reaction.  The general theory of rate presented
in the previous section assumes always
non-Markovian. On the other hand, the experimental
observations appear often Markovian.
It is desirable to have a working criterion
on the nature of ET kinetics.
 In contact with experiments,
let us consider the scaled population,
\be \label{P_r}
  \Delta(t)\equiv \frac{P_{j}(t)-P_{j}(\infty)}
  {P_{j}(0)-P_{j}(\infty)} ;   \ \ \ j=a, b.
\ee
It does not depend on the state-index due to
the identity of $P_a(t)+P_b(t)=1$.
The simple (Markovian) kinetic rate equation,
$\dot P_a(t)=-kP_a(t)+k'P_b(t)$,
can be represented in terms of the
scaled population as
\be\label{DelMar}
  \dot\Delta_{\rm Mar}(t)=-w\Delta_{\rm Mar}(t).
\ee
The involving decay constant, $w=k+k'$,
 could be measured readily, if the rate does behave like Markovian.
The forward and backward reaction rate constants
can then be evaluated as
$k=wK_{\rm eq}/(1+K_{\rm eq})$ and $k'=w/(1+K_{\rm eq})$,
respectively.

 A non-Markovian rate process can be described
by the {\it kinetic rate} memory kernel $\hat w(t-\tau)$
via
\be\label{hatw}
  \dot\Delta(t) \equiv -\int_0^{t}\!\!d\tau\,
   \hat w(t-\tau)\Delta(\tau) .
\ee
 In other words, one can deduce the rate kernel from the
population evolution as
$\hat w(t) =L^{-1}\{[1-s\ti\Delta(s)]/\ti\Delta(s)\}$.
Here $\ti\Delta(s)$ denotes the Laplace transform
of $\Delta(t)$, while $L^{-1}\{\cdot\}$
the inverse Laplace transform,
i.e.
\be\label{kinrateS}
  w(s) \equiv L\{\hat w(t)\}
  \equiv \int_0^{\infty}\!\!dt\, e^{-st}\hat w(t)
  = \frac{1-s\ti\Delta(s)}{\ti\Delta(s)}.
\ee
 In \App{thapp_Laplace}, we present the
 explicit expression of $w(s)$ in terms of
 $k(s)$ and $k'(s)$; see \Eq{wkkps},
 together with some useful identities  in relation
 to the Laplace transform. Unlike $w(s)$,
 the population evolution alone is in general not sufficient to
 determine both $k(s)$ and $k'(s)$.

 Equivalent to the kinetic rate kernel $\hat w(t)$ defined
 in \Eq{hatw}, one may also define
 the time-local rate $W(t)$ via $\dot\Delta(t)=-W(t)\Delta(t)$.
Note that while the kernel $\hat w(t)$
is always well behaved, its time-local counterpart,
$W(t)=-\dot\Delta(t)/\Delta(t)$, may diverge
at certain time, say $t'$, due to the possibility of
$\Delta(t')=0$ in an underdamped non-Markovian rate process.
At time $t'$, the effective kinetics reduces to the zero-order
rate process, since $W(t')\Delta(t')$ remains finite.
Moreover, \Eq{hatw} implies also
that $W(t\!=\!0)=0$ and $W(t\rightarrow\infty)=w(s\!=\!0)\equiv w_0$.
The latter is nothing but the fact that
long-time regime is Markovian.

 To determine the short-time behavior for the population dynamics,
 we notice the identity,
\be\label{hatK0}
 \hat k(t\!=\!0) = \hat k'(t\!=\!0) = 2V^2/\hbar^2.
\ee
This can be obtained directly by considering the fact that
the short-time  behavior is identical to
$\hat k_{\NA}(t\!=\!0)$ (cf.\ \Eq{kNAall}),
regardless whether the reaction is nonadiabatic or not.
Together with \Eq{ksct0}, and setting $P_a(0)=1$, we obtain
\be \label{wt0}
 \hat w\equiv \hat w(t\!=\!0) = \frac{2V^2}{\hbar^2}
  \frac{1+K_{\rm eq}}{K_{\rm eq}}.
\ee
It determines the short-time behavior
of the scaled population dynamics, as
$\dot\Delta(0)=0$ and $\ddot\Delta(0)=-\hat w$
that are implied in \Eq{hatw}.

 The short- and long-time behavior described above
can be summarized as
\be\label{asympDelta}
  \Delta(t\rightarrow 0) \approx e^{-\hat wt^2/2},
\ \ \
  \Delta(t\rightarrow\infty) \approx e^{-wt}.
\ee
 Besides the above asymptotic behaviors,
$\Delta(t)$ is in general also influenced by the coherent motion
or quantum beat, as long as the rate process
goes beyond the second-order in the transfer coupling $V$.
 As the asymptotic behaviors are concerned,
it may suggest to use the Kubo's line shape
function,\cite{Kub69101}
\be\label{Delc}
  \Delta_{\rm K}(t)= \exp\left[
    -\kappa^{-2}(e^{-\hat wt/w}-1+\hat wt/w)
   \right],
\ee
to analyze the nature of population transfer dynamics.
The involving Kubo's Markovianicity parameter
is (cf.\ \Eqs{wt0} and \ref{asympDelta})
\be\label{kappa}
 \kappa \equiv \frac{\hat w^{1/2}}{w}
  = \left(\frac{2K_{\rm eq}}{1+K_{\rm eq}}\right)^{1/2}
   \frac{V}{\hbar k}.
\ee
The rate process assumes Markovian or non-Markovian,
when $\kappa>1$ or $\kappa<1$, respectively.

\section{Effects of quantum solvation: Numerical results and discussions}
\label{thnum1}

\subsection{Debye solvent model and general remarks}
\label{thnumA}

  For the numerical demonstrations presented hereafter,
 we consider the ET system
 in \Eq{HT} or \Fig{fig1}, with the solvation
 correlation function in \Eq{Ct} being characterized by
 \be \label{DebyeC}
  C(t) =  \eta \exp(-t/\tau_{\rm L}).
 \ee
 This is the Debye solvent model,
 with the longitudinal relaxation time $\tau_{\rm L}$
 and the pre-exponential parameter\cite{Tan89101,Han0611438}
 \be \label{eta}
   \eta = \lambda(2\kT - i\hbar/\tau_{\rm L}).
 \ee
 For the above ET system, the analytical expressions
 for the reduced density matrix $\rho(t)$ and
 the involving dynamics rate functions, $k(s)$ and $k'(s)$, have
 all been constructed in Ref.\ \onlinecite{Han0611438}.
 Note that \Eq{DebyeC} satisfies a semiclassical
 fluctuation-dissipation theorem, valid when
 the temperature is comparable with or higher than the system's
 transition energy.\cite{Han0611438,Tan06082001}
 This is the only approximation involved in this paper.

   To elucidate the quantum solvation effect on ET,
 especially as the reaction mechanism is concerned,
 the rate in the classical solvation limit will also be
 evaluated as a reference. The classical solvation
 correlation function assumes real;
 i.e., $\eta\rightarrow\eta_{\rm cl}=2\lambda\kT$
 when $\kT/\hbar\gg \tau^{-1}_{\rm L}$.
 It follows $|\eta-\eta_{\rm cl}|/\eta_{\rm cl} =
 0.5 \tau_{\rm ther}/\tau_{\rm L}$, with
 $\tau_{\rm ther} \equiv \hbar/(\kT)$ denoting
 the {\it thermal time} ($\tau_{\rm ther}=26$\,fs for $T=298$ K).
 The quantum solvation effect can be significant
 when $\tau_{\rm L}<\tau_{\rm ther}$.

  It is also noticed that the solvation longitudinal
 relaxation time $\tau_{\rm L}$ is proportional to the solvent
 viscosity.\cite{Yan884842,Yan896991}
 In this sense, the $k$ versus $\tau_{\rm L}$ behavior
 can be referred as the {\it rate--viscosity} character.
 This connection suggests the well-established
 Kramers' picture of the solvation effect on
 chemical reaction\cite{Kra40284,Han90251}
 be exploited in the following to elucidate the
 underlying ET reaction mechanism.

\subsection{Quantum vs.\ classical solvation effects}
\label{thnumB}
   As the mechanism is concerned,
 the effect of quantum versus classical solvation is expected
 to be most distinct in the following two scenarios (cf.\ \Fig{fig1}).
 One is the symmetric ET system ($E^{\circ}=0$) in which
 the Fermi resonance enhanced quantum tunneling
  is anticipated.
 Another is the classical barrierless system
 ($E^{\circ}+\lambda=0$) where the Marcus'
  inversion takes place.
 Figure \ref{fig2} depicts the evaluated rate-viscosity characteristics
 for these two scenarios of ET system,
 at two values of transfer coupling:
 $V=1$\,kJ/mol (upper-panels) and
 $V=0.01$\,kJ/mol (lower-panels).
 The solvation energy $\lambda=3$ kJ/mol and  temperature $T=298$ K.

 In the high-viscosity ($\tau_{\rm L}>\tau_{\rm ther}$)  regime,
 the difference between the quantum (solid-curve)
 and the classical (dashed-curve) in each panel
 diminishes (at the qualitative level).
 This observation is consistent with
 the physical picture that the high viscosity
 (or slow motion) implies a large effective mass
 and thus leads to the classical solvation limit.
 Observed here is also the Kramers' fall-off
 behavior\cite{Kra40284,Han90251}
 for the adiabatic rates in the upper panels.
 This is the diffusion limit; the higher the solvent
 viscosity is, the more backscattering (or barrier re-crossing
 in the classical sense) events will be.
 For the nonadiabatic processes in the lower panels,
 the backscattering effects are quenched.
 This accounts for the plateau, observed in the lower-panel (c) or (d),
 at which the Marcus' nonadiabatic ET regime is reached,
 and the rate becomes independent of viscosity.

   In the low-viscosity ($\tau_{\rm L}<\tau_{\rm ther}$) regime,
 the difference between quantum and classical solvation is at
 the mechanism level for the two specified types of ET systems.
 For the symmetric ($E^{\circ}=0$) system
 [the left-panel (a) or (c) of \Fig{fig2}],
 the quantum rates (solid-curves)
 are apparently Fermi resonance-assisted tunneling dominated
 processes, while the classical rates (dash-curves)
 are barrier crossing events.
 In contrast to that $k \gg k_{\rm cl}$
 in the $\tau_{\rm L}<\tau_{\rm ther}$ regime
 for the symmetric ($E^{\circ}=0$) system,
 observed is the opposite result of $k \ll k_{\rm cl}$
 for the Marcus' barrierless
  ($E^{\circ}+\lambda=0$) system
 [the right-panel (b) or (d) of \Fig{fig2}].
  In the latter case, the classical rate
 does behave as a barrierless reaction,
 but the quantum rate exhibits the Kramers'
 barrier-crossing characteristics as a function of viscosity.
  This may indicate that for the ET in the classical barrierless system
 there is an effective viscosity-dependent barrier that
 vanishes as $\tau_{\rm L}$ increases.
 In other words, the barrier of $(E^{\circ}+\lambda)^2/(4\lambda)$,
 as depicted in the ET schematics \Fig{fig1}, is the static picture
 that assumes the solvation potential being fixed.
 This picture is valid in the high-viscosity (or slow-modulation)
 regime, but no longer true in the low-viscosity
 (or fast-modulation) regime. In general,
 the effective solvation potential for ET reaction
 is solvent viscosity dependent.

  To confirm the above observed ET mechanism-related features,
 the values of $\lambda$ and $T$ are varied, and the results for
 $V=1$\,kJ/mol are summarized in \Fig{fig3}.
 Here, $k/k_{\rm cl}$ as functions of $\tau_{\rm L}/\tau_{\rm ther}$
 are depicted. This figure verifies that $\tau_{\rm L}/\tau_{\rm ther}$ does
 serve a proper measure for the nature of solvation.
 The reaction mechanism turnover occurs
 at $\tau_{\rm L}= \tau_{\rm ther}$,
 for the symmetric ($E^{\circ}=0$) case,
 and classical barrierless ($E^{\circ}+\lambda=0$)
 ET systems. In the former case, it is
 changed from the tunneling
 to barrier-crossing, while in the latter case from the
 barrier-crossing to barrierless ET rate process.

 Let us now consider the Marcus-type inversion behaviors.
 Figure \ref{fig4} presents the Marcus-plots,
 the logarithmic rate log $k$ versus reaction
 endothermicity $E^{\circ}$, in relation to \Fig{fig2}.
 Three values of $\tau_{\rm L}/\tau_{\rm ther}$ are chosen as
 0.1 (solid-curves), 1 (dot-curves), and 10 (dash-curves),
 to represent the low-, intermediate-,
 and high-viscosity regimes, respectively.

  Note that $E^{\circ}=-\lambda$ does
 represent the classical barrierless ET systems
 in all cases in study. All the classical rates
 [right-panels, \Fig{fig4}(b) and (d)]
 have inversions occurring at $E^{\circ}=-\lambda$.
 Moreover, all of them are symmetric
 about $E^{\circ}=-\lambda$.
 The Marcus' parabolic character is recovered
 in the high-viscosity, nonadiabatic and classical limit,
 and is practically the dashed curve in \Fig{fig4}(d).

 The quantum rates depicted in the left-panels
 [\Fig{fig4}(a) and (c)] also show inversion
 behavior. However, the observed inversion region depends sensitively
 on the solvent viscosity, and shows also an asymmetric character
 that will be elaborated soon.
 Apparently, the quantum rate inversion region
 is closely related to the interplay between the
 barrier-crossing and tunneling processes.

 Let us start with \Fig{fig4}(a).
 Consider first the high-viscosity regime, in which,
 according to the analysis made earlier for \Fig{fig2},
 the ET process is qualitatively the same as its classical
 counterpart. Consequently the dash-curve in \Fig{fig4}(a)
 has the inversion region around the classical barrierless
 position at $E^{\circ}=-\lambda$.
 Consider now the low-viscosity regime, in which, again,
 according to the analysis earlier
 there is always a nonzero barrier for the ET reaction,
 covering over the entire range of $E^{\circ}$
 including the value of $E^{\circ}=-\lambda$.
 This explains the inversion behavior of
 the solid curve of \Fig{fig4}(a) that is peaked
 at the resonant position of $E^{\circ}=0$.
 As the viscosity increases, the inversion region
 smoothly shifts from the resonant peak position
 $E^{\circ}=0$ to the classical barrierless position
 of $E^{\circ}=-\lambda$.

 To explain the asymmetric property of the
 quantum inversion behavior as depicted
 in the left panels of \Fig{fig4}, recall
 that $k(-E^{\circ})\approx k'(E^{\circ})$, the backward reaction
 rate, and $k(E^{\circ})<k'(E^{\circ})$ for an endothermic
 ($E^{\circ}>0$) reaction. This leads immediately
 to the asymmetric property of the solid-curve
 in \Fig{fig4}(a) or (c), in which
 the blue (endothermic) wing falls off faster
 than its red (exothermic) wing.
 This asymmetry decreases as the viscosity increases,
 since the high viscosity regime behaves classically.

\section{The adiabatic and Markovian characters of the reaction}
\label{thnum2}

  We are now in the position to demonstrate
 (for the cases of quantum solvation only)
 the issues in relation to the nature of ET rate process,
 as addressed in \Sec{thsec3}.
  Let us start with the adiabatic/nonadiabatic character
 depicted in \Fig{fig5}.
  Here,  each individual $k$ (thin)-curves,
 as function of $\tau_{\rm L}$, is decomposed into its
 adiabatic $k_{\mbox{\tiny A}}$
 and nonadiabatic $k_{\NA}$ components (\Eq{kA_def}).
 These two components are given in the upper-panel
 [\Fig{fig5}(a) or (b)]
 and the lower-panel [\Fig{fig5}(c) or (d)], respectively.
 The left-panels are for the $E^{\circ}=0$ system
[\Fig{fig5}(a) and (c)], while
 the right-panels are for the $E^{\circ}+\lambda=0$
 system [\Fig{fig5}(b) and (d)].
 In each panel, four values of transfer coupling  are used:
 $V/$(kJ/mol) = 0.25 (dotted), 0.5 (dashed), 1 (solid),
 and 2 (dash-dotted).

   It is observed that in the low-viscosity
 ($\tau_{\rm L}<\tau_{\rm ther}$) regime,
 the reaction is nonadiabatic; see \Fig{fig5}(c) and (d).
  This observation may be understood as follows.
   In the low-viscosity regime, the solvent fluctuates fast
 and stabilizes the ET system in the acceptor state
 before backscattering taking place.
 As results, the reaction is nonadiabatic in the
 low-viscosity regime, at least for the
 range of transfer coupling strength considered
 here.
  The above picture is also consistent with
  the observation that in the diffusion
 ($\tau_{\rm L}\gg\tau_{\rm ther}$) limit,
 where $k\propto 1/\tau_{\rm L}$,
 the reaction assumes an adiabatic rate process;
 see \Fig{fig5}(a) and (b).

  Figure \ref{fig6} presents the adiabaticity parameters,
 $k_{\NA}/k_{\mbox{\tiny A}}$, as function of
 reaction endothermicity $E^{\circ}$, at $V=1$\, kJ/mol,
 with the three specified values of
 $\tau_{\rm L}/\tau_{\rm ther}=0.1$, 1, and 10.
 The inverse of these values are also used individually to scale
 the corresponding adiabaticity curves, as depicted in \Fig{fig6}.
 The adiabaticity curve in
 high-viscosity ($\tau_{\rm L}/\tau_{\rm ther}=10$) regime
 looks all normal, being of the minimum about where
 the rate maximum is;
 cf.\ the dash-curve in \Fig{fig4}(a).
%
 The faster the ET passage, the less adiabatic
 (or more surface-hopping) the reaction would be.
 This feature remains largely unchanged
 in the intermediate-viscosity
 ($\tau_{\rm L}/\tau_{\rm ther}=1$) case,
 except for that the adiabaticity minimum now
 is slightly {\it negative}, or $k>k_{\NA}$.
 It contradicts with the argument made earlier for \Eq{kA_def}.
 The abnormality here may be accounted for, at least partially,
 by the associated tunneling rate process, as elaborated below.

  The abnormality in the adiabaticity parameter
 is most striking around the symmetric ($E^{\circ}\approx 0$)
  system in the low-viscosity ($\tau_{\rm L}/\tau_{\rm ther}=0.1$)
 regime. This is tunneling dominant scenario, as discussed
 in \Sec{thnum1}. Considered here is also the strong transfer coupling
 case of $V=1$\,kJ/mol.
 Thus, the observed abnormality is most likely caused by
 the coherent ET reaction.
 The argument for $k<k_{\NA}$, due to the backscattering-induced
 total rate reduction, may no longer be valid.
 Moreover, we will see soon that the ET reaction
 in the observed abnormal region is non-Markovian,
 leading to the rate constant description
 inadequate at all; see \Fig{fig10} and the comments there.

  We now turn to the Markovian/non-Markovian nature of ET reaction.
 The key quantity here is the Markovianicity parameter
 $\kappa$ (\Eq{kappa}). It involves the reaction rate $k$
 and the equilibrium constant $K_{\rm eq}$.
   The detailed knowledge on how $K_{\rm eq}$
 depends on the Debye solvent parameters can be found
 in Ref.\ \onlinecite{Han0611438}.
 For the rate in the low-viscosity regime,
 $k \approx k_{\NA}$, as depicted in the lower-panels of \Fig{fig5},
 resulting in $k\approx 2V^2\Gamma/(E^{\circ 2}+\Gamma^2)$, with
 $\Gamma=2\lambda\kT\tau_{\rm L}$ for the Debye solvent
 model in study;  cf.\ \Eq{kNAall} with \Eq{DebyeC}.
 Thus,
 \be \label{kappappr}
  \kappa\approx \lambda\kT\tau_{\rm L}/V;
\ \ {\rm when\ }E^{\circ}=0 \
  {\rm and\ }\tau_{\rm L} < \tau_{\rm ther}.
 \ee
 This equation can be used to estimate the Markovianicity parameter
 for a symmetric ET system
  ($E^{\circ}=0$, implying also $K_{\rm eq}=1$)
 in the low-viscosity regime.

   Figure \ref{fig7} shows the evaluated
 Markovianicity $\kappa$ (\Eq{kappa}) as function of
 $\tau_{\rm L}/\tau_{\rm ther}$,
 for the same ET systems as \Fig{fig2} (without the classical
  solvation parts).
 For $E^{\circ}+\lambda=0$ (lower-panel),
 the ET rate process behaves Markovian ($\kappa>1$), even for
 the strong transfer coupling ($V =1$\,kJ/mol) case.
  Apparently, the non-Markovian ($\kappa<1$) rate process
 is most likely to occur in the Fermi-resonance tunneling regime,
 where $E^{\circ}=0$ and $\tau_{\rm L} < \tau_{\rm ther}$,
 with the approximated expression of $\kappa$ given in \Eq{kappappr}.
 The above comments are further confirmed by \Fig{fig8},
 in which the Markovianicity $\kappa$ is plotted
 as  function of $E^{\circ}$, at the three representing values
 of $\tau_{\rm L}/\tau_{\rm ther}$.

    Figure \ref{fig9} depicts the scaled population evolution,
  $\Delta(t)$ (\Eq{P_r}), for the ET systems of $E^{\circ}=0$
 (three left-panels) and $E^{\circ}+\lambda=0$ (three right-panels).
   The upper [(a) and (b)], middle [(c) and (d)],
 and lower [(e) and (f)] panels are of
 the specified values of viscosity,
 $\tau_{\rm L}/\tau_{\rm ther}=0.1$, 1 and 10,
 respectively.
  The transfer coupling strength is $V=1$ kJ/mol.
  The weak transfer coupling ($V=0.01$\,kJ/mol) counterparts,
 as depicted in the inserts of individual panels,
  are shown all Markovian, with $\kappa\gg 1$
  for their values of Markovianicity; cf.\ \Fig{fig7} and \Fig{fig8}.
  Included for comparison in each panel are also the Kubo's
  $\Delta_{\rm K}(t)$ (\Eq{Delc}) and
  Markovian $\Delta_{\rm Mar}=\exp(-wt)$
  counterparts, where $w=k+k'$.
    As the envelop of population evolution is concerned,
    the significant non-Markovian nature
  is only observed in \Fig{fig9}(a), with
 the relevant part is enlarged in \Fig{fig10}.
   This is a symmetric ET system in
 the strong transfer ($V=1$\,kJ/mol) coupling
 and low-viscosity ($\tau_{\rm L}/\tau_{\rm ther}=0.1$) regime.
 The corresponding Markovianicity value ($\kappa=0.3$) is found to
 agree well with the aforementioned approximate expression
 of \Eq{kappappr}.

\section{Concluding remarks}
\label{thsum}

   The quantum solvation, adiabatic/nonadiabatic
 and Markovian/non-Markovian characters are
 important issues in understanding chemical
 reaction, including ET in solution.
  Here, we have revisited these issues
 in a unified and transparent manner,
 with the aid of Debye solvent model (\Eq{DebyeC})
 that supports an analytical solution without
 additional approximations.\cite{Han0611438}
  The physical picture discussed in this work
 is however rather general.

  We have presented a full account for
 the effect of quantum solvation on the
 ET rate process (\Sec{thnum1}).
   Not just can it change
 a barrier-crossing event to tunnelling,
 the quantum nature of solvent can also lead
 a classical barrierless reaction
 to an effective barrier-crossing rate process.
 The resulting rate may differ from its classical
 counterpart by order of magnitude.
  The quantum solvation is found to be distinctly important
 in low-viscosity (fast-modulation) solvents.
  For a realistic solvent that consists of
 multiple correlation time scales,
 only the slow-modulation solvent modes can be treated
 classically.

  The adiabatic--nonadiabatic decomposition of rate (\Eq{kA_def})
 is practically useful, provided  that the total rate
 $k$ can be experimentally measured and the nonadiabatic
 rate $k_{\NA}$ can be readily evaluated via \Eq{kNAall}.
 The adiabaticity parameter $k_{\NA}/k_{\mbox{\tiny A}}= k_{\NA}/k-1$,
 as inferred from \Eq{kA_def}, can then be used to
 discuss the adiabatic/nonadiabatic nature of reaction.
 Interestingly, a negative adiabaticity may indicate
 there is a certain degree of quantum
 tunneling taking place; cf.\ \Fig{fig6} and its comments.

  We have also proposed to use the
 Markovianicity parameter $\kappa$ (\Eq{kappa}),
 based on the Kubo's motional narrowing function,
 for analyzing
 the Markovian/non-Markovian nature of electron transfer
 rate process. Note the solvent relaxation time scale
 is typically in the order of picosecond,
 while $\tau_{\rm ther}$ for room temperature is 26 fs.
 This amounts to the high-viscosity or slow-modulation
 regime of present studies on ET.
 The resulting Markovianicity parameter,
 as depicted in \Fig{fig7} and \Fig{fig8},
 is typically of $\kappa>1$ for the aforementioned
 typical cases. This may account for
 why most experimental observations do support
 the Markovian ($\kappa>1$) rate constant description.

   We have pointed out that a non-Markovian rate
  process is most likely to occur in the
  symmetric ET system at the fast-modulation
  regime.  It is just the opposite to the spectroscopic case.
  According to the motional narrowing picture,
  the fast modulation leads to a Markovian spectroscopic
  process.\cite{Kub66255, Kub69101}
  The above seemingly counter-intuitive phenomenon
  in relation to the nature of rate process may
  be understood as follows.
  First of all, the motional narrowing picture is applicable
  to the spectrum of rate kernel, rather than the population
  evolution itself. The narrower the rate kernel spectrum is,
  the less Markovian of rate process
  would be. However, this is not
  the complete picture.
  The peak position of the rate kernel spectrum,
  in relation to where rate constant is evaluated,
  should also be considered.
  It shifts from the classical Marcus' inversion
  position at $E^{\circ}=-\lambda$ in the slow-modulation limit,
  to the quantum resonant tunneling at
  $E^{\circ}=0$ in the fast-modulation (or motional-narrowing) regime,
  see \Fig{fig4} (left-panels).
 Equation (\ref{kappappr}) that is achieved at
 $E^{\circ}=0$ can be considered as the lower bound
 of the Markovianicity $\kappa$ for the ET rate process.
 in the fast modulation regime.
  In this regime, the population transfer may also
 exhibit the quantum beat feature that is
 non-Markovian in a strict sense.
 We shall investigate these complex cases elsewhere.

\begin{acknowledgments}
 Support from the RGC Hong Kong (604006), NNSF of
 China (50121202), and
 National Basic Research Program of China
 (2006CB922004) is acknowledged.
 R.\ X.\ Xu would also like to thank the support from
 the NNSF of China (20403016 and 20533060)
 and Ministry of Education of China (NCET-05-0546).
\end{acknowledgments}

\appendix

\section{Perturbative rate kernels}
\label{thapp_perturate}
   In this appendix, we shall treat the nonadiabatic rate problem,
 on the basis of the standard perturbation theory on the
 reduced density matrix $\rho(t)$,
  assuming the transfer coupling last term of \Eq{HT}, is weak.
 At the initial time $t=0$, the total composite density
 operator is $\rho_{\rm T}(0) = \rho_{a}^{\rm eq}|a\ra\la a|$.
   Consider \Eq{rhosmat} for the coherence components, which
 for the two-level ET system reads explicitly as
 (setting $\hbar=1$ in this appendix)
 \be\label{coherence}
  \alpha\ti\rho_{ba} =-iV(\ti P_a-\ti P_b)
  - z\ti P_b - y\ti\rho^{\ast}_{ba}.
 \ee
 Here $\alpha=s+i(E^{\circ}+\lambda)+x$, and
 \be
   x\equiv \Pi_{ba,ba}, \ \ y\equiv \Pi_{ba,ab}, \ \ z\equiv \Pi_{ba,bb}.
 \ee
 Combining the initial condition $P_a(t=0)=1$ and the perturbative
 action of $\hat V$ together result immediately
 in $\ti P_a^{\{0\}}=1/s$ and
\[
  0=\ti P_a^{\{2k+1\}}=\rho_{ba}^{\{2k\}}
   = \alpha^{\{2k+1\}} = y^{\{0\}} = y^{\{2k+1\}}=z^{\{2k\}}.
\]
 Therefore, the lowest order in \Eq{coherence} reads
\be\label{srho1_1}
 \alpha^{\{0\}}\ti\rho^{\{1\}}_{ba}=-iV/s.
\ee
 On the other hand, the standard first-order
perturbative expression is
\be\label{rho1t}
 \rho^{(1)}_{ba}(t)= -iV\int_0^{t}\!d\tau
    \exp[-g(\tau)].
\ee
 Here
 \be\label{app_G0}
  \exp[-g(t)] =
    \left\la \exp_{+}\left\{-i\int_0^t\!d\tau\,
      [E^{\circ}+U(\tau)]\right\} \right\ra .
 \ee
 Using the second-cumulant expansion expression, which is
 exact for the Gaussian solvation process, results in
 \be\label{app_gt}
   g(t) = i(E^{\circ}+\lambda)t +
   \int_{0}^t\!\!d\tau\int_0^{\tau}\!\!d\tau'
    C(\tau').
 \ee
We have then
\be\label{srho1_2}
 s\rho^{\{1\}}_{ba}(s)=-iV L\{\exp[-g(t)]\} \equiv -iVJ(s).
\ee
Together with \Eq{srho1_1}, we obtain
\be\label{alp0}
  \alpha^{\{0\}}(s)=1/J(s).
\ee
 Together with $y^{\{0\}}=0$, \Eq{Ks} to the lowest order
 reads $k_{\NA}(s) = (2V^2/\hbar^2) {\rm Re} J(s)$,
 where $J(s)$ denotes the Laplace transform of $\exp[-g(t)]$
 (cf.\,\Eq{srho1_2}). We obtain therefore \Eq{kNAall}.

\section{Some useful relations for rates}
\label{thapp_Laplace}

 Let us first present some basic relations in connection to
the Laplace transform, defined for $s \geq 0$ as
 \be\label{Lapdef}
    \ti f(s) \equiv L\{f(t)\} \equiv\int_0^\infty\!\!dt\,e^{-st}f(t).
 \ee
 It satisfies the boundary condition,
 $L\{\cdot\}\big|_{s\rightarrow\infty}=0$,
 and
 \be\label{Lf1f2}
   L\left\{\int_0^t\!d\tau f_1(t-\tau)f_2(\tau)\right\}
  =\ti f_1(s)\ti f_2(s),
 \ee
 \be\label{Ldotf}
   L\{\dot f(t)\} = s\ti f(s) -f(0).
 \ee
 Using \Eq{Ldotf}, together with the identities
 of $L\{\cdot\}\big|_{s\rightarrow\infty}=0$
 and $L\{\dot f(t)\}|_{s=0} =f(\infty)-f(0)$,
 we obtain immediately
 \be\label{f0infty}
  f(0)=\lim_{s\rightarrow\infty}[s\ti f(s)],
 \ \ \
  f(\infty)=\lim_{s\rightarrow 0}[s\ti f(s)].
 \ee

 We are now in the position to derive
some useful relations between the non-Markovian rate variables
appearing in \Sec{thsec2}.
From \Eq{cop} and $P_a(t)+P_b(t)=1$, we have
\be\label{tilPas1}
  \tilde P_a(s)=\frac{P_a(0)+k'(s)/s}{s+k(s)+k'(s)}.
\ee
From \Eqs{P_r} and \ref{kinrateS}, we have
 \be \label{tilPas2}
  \tilde\Delta(s)=\frac{\tilde P_a(s)-P_a(\infty)/s}{P_a(0)-P_a(\infty)}
  = \frac{1}{s+w(s)}.
 \ee
The above two equations lead to
 \be\label{wkkps}
        w(s)=\frac{s[P_a(0)k(s)-P_b(0)k'(s)]}
     {s[P_a(0)-P_a(\infty)]-P_a(\infty)k(s)+P_b(\infty)k'(s)}\,.
 \ee
Together with the first identity of \Eq{f0infty}, we have
 \be\label{ksct0}
   \hat w(t=0) =
  \frac{P_a(0)\hat k(t=0)-P_b(0)\hat k'(t=0)}  {P_a(0)-P_a(\infty)}.
\ee
The above relation will be used in deriving \Eq{wt0}.


\clearpage

\begin{figure}
\caption{
 Schematics of solvent potentials $V_a$ and $V_b$
 for the ET system in the donor and acceptor states,
 respectively, as functions of the solvation coordinate
 $U\equiv h_b-h_a=V_b-V_a-E^{\circ}$, with
  $E^{\circ}$ being the ET endothermicity
 and $\lambda = \la U\ra$ the solvation energy.
 The classical barrierless system is that of $E^{\circ}+\lambda=0$.
}
\label{fig1}
\end{figure}

 \begin{figure}
 \caption{
   Electron transfer rate $k$ (solid-curves),
  as function of solvent longitudinal
   relaxation time $\tau_{\rm L}$, with $\lambda=3$\,kJ/mol
   at $T=298$\,K.
   Left-panels (a) and (c): $E^{\circ}=0$;
   right-panels (b) and (d): $E^{\circ}+\lambda=0$;
   upper-panels (a) and (b): $V=1$ kJ/mol;
   lower-panels (c) and (d): $V=0.01$ kJ/mol.
  Included in each panel is also the classical solvation
  counterpart $k_{\rm cl}$ (dash-curves).
    Note that $\tau_{\rm ther}\equiv\hbar/(\kT)= 10^{-1.6}$ps.
}
 \label{fig2}
 \end{figure}

\begin{figure}
 \caption{
 The ratio of quantum versus classical rates, $k/k_{\rm cl}$,
  as function of $\tau_{\rm L}/\tau_{\rm ther}$,
 with the transfer coupling $V=1$ kJ/mol,
 at various specified values of $\lambda$ and $T$.
}
 \label{fig3}
 \end{figure}

\begin{figure}
 \caption{
 The rate $k$ (left-panels) and the classical counterpart
 $k_{\rm cl}$ (right-panels), as function of $E^{\circ}$,
 with $\lambda=3$ kJ/mol at $T=298$ K;
 upper-panels (a) and (b): $V=1$ kJ/mol;
 lower-panels (c) and (d): $V=0.01$ kJ/mol.
 The three values of relative relaxation time scale,
 $\tau_{\rm L}/\tau_{\rm ther}=$ 0.1 (solid), 1 (dotted), and 10 (dashed)
  are chosen to represent the low, intermediate,
 and high-viscosity regimes, respectively.
}
 \label{fig4}
 \end{figure}

\begin{figure}
 \caption{The decomposition of rate $k$ (thin-curves),
  following \Eq{kA_def}, into the
 adiabatic $k_{\mbox{\tiny A}}$ (upper-panels)
 and nonadiabatic $k_{\NA}$ (lower-panels) components,
 and plotted as functions of $\tau_{\rm L}/\tau_{\rm ther}$,
  with $\lambda=3$\,kJ/mol and $T=$\,298K.
 Left-panels (a) and (c): $E^{\circ}=0$;
 right-panels (b) and (d): $E^{\circ}+\lambda=0$.
 Each panel involves four values of transfer coupling strength:
  $V/$(kJ/mol) = 0.25 (dotted), 0.5 (dashed), 1 (solid),
  and  2 (dash-dotted), respectively.
   }
 \label{fig5}
 \end{figure}

\begin{figure}
 \caption{
   The adiabaticity $k_{\NA}/k_{\mbox{\tiny A}}$ as
 function of $E^{\circ}$, for
 $\tau_{\rm L}/\tau_{\rm ther}=$ 10, 1 and 0.1.
 The inverse of $\tau_{\rm L}/\tau_{\rm ther}$
 is also used to the scale individual adiabaticity curve.
 $V=1$\,kJ/mol, $\lambda=3$\,kJ/mol, and $T=$\,298K.
   }
 \label{fig6}
 \end{figure}

\begin{figure}
 \caption{
 The Markovianicity parameter $\kappa$ (\Eq{kappa}),
  as function of $\tau_{\rm L}/\tau_{\rm ther}$,
 at $V=1$ kJ/mol (solid) and 0.01 kJ/mol (dashed),
 with $\lambda=3$\,kJ/mol and $T=298$\,K,
 for (a) $E^{\circ}=0$ and (b)   $E^{\circ}+\lambda=0$.
}
 \label{fig7}
 \end{figure}

\begin{figure}
 \caption{
  The Markovianicity parameter $\kappa$ (\Eq{kappa})
  as the function of $E^{\circ}$,
 with $\lambda=3$ kJ/mol and $T=298$ K,
 at $\tau_{\rm L}/\tau_{\rm ther}=$ 0.1 (solid),
 1 (dotted) and 10 (dashed):
 (a) $V=1$ kJ/mol and (b) $V=0.01$ kJ/mol.
 }
\label{fig8}
\end{figure}

\begin{figure}
 \caption{The scaled population $\Delta(t)$ (\Eq{P_r}) evolution,
 evaluated at $\tau_{\rm L}/\tau_{\rm ther}=$0.1 (upper-panels),
 1 (middle-panels), and 10 (lower-panels);
 Left-panels: $E^{\circ}=0$;
 right-panels: $E^{\circ}+\lambda=0$.
 Here, $V=1$\,kJ/mol, $\lambda=3$\,kJ/mol, and $T=298$\,K.
 Included in each panel are also the corresponding
 Kubo's $\Delta_{\rm K}(t)$ (\Eq{Delc}; dashed)
 and Markovian $\Delta_{\rm Mar}(t)$ (dotted).
 The insert in each panel is the $V=0.01$\,kJ/mol counterpart,
 where all these three curves are identical.
}
 \label{fig9}
 \end{figure}

\begin{figure}
 \caption{ The amplified portion of \Fig{fig9}(a). The
  scaled population evolution $\Delta(t)$ (solid),
 Kubo's $\Delta_{\rm K}(t)$ (dashed),
 and Markovian $\Delta_{\rm Mar}(t)$ (dotted).
}
 \label{fig10}
 \end{figure}

\clearpage
\begin{center}
\centerline{\includegraphics[width= 0.95\columnwidth,angle=0]{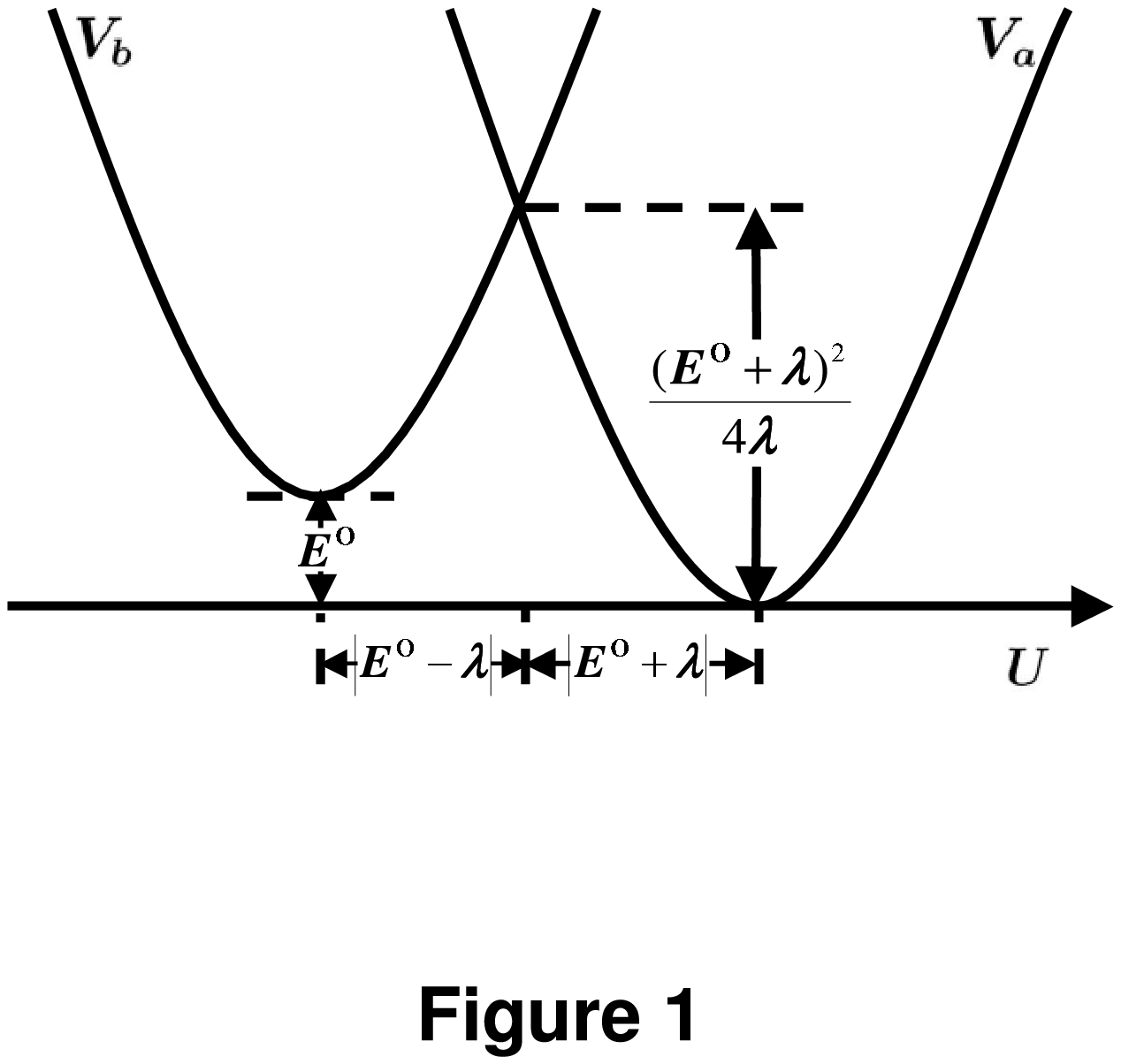}}
\centerline{\includegraphics[width= 0.90\columnwidth,angle=0]{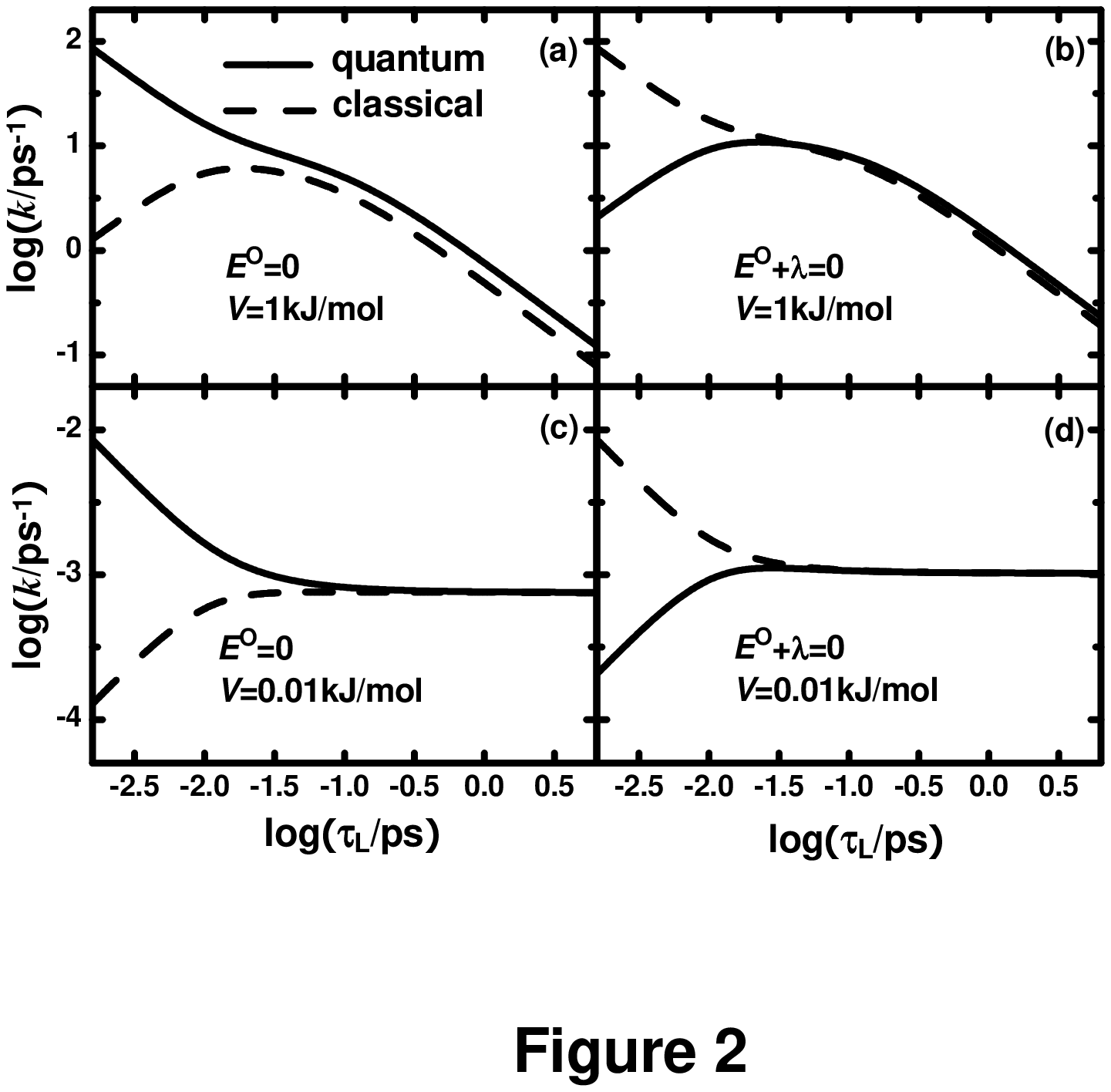}}
\centerline{\includegraphics[width= 0.90\columnwidth,angle=0]{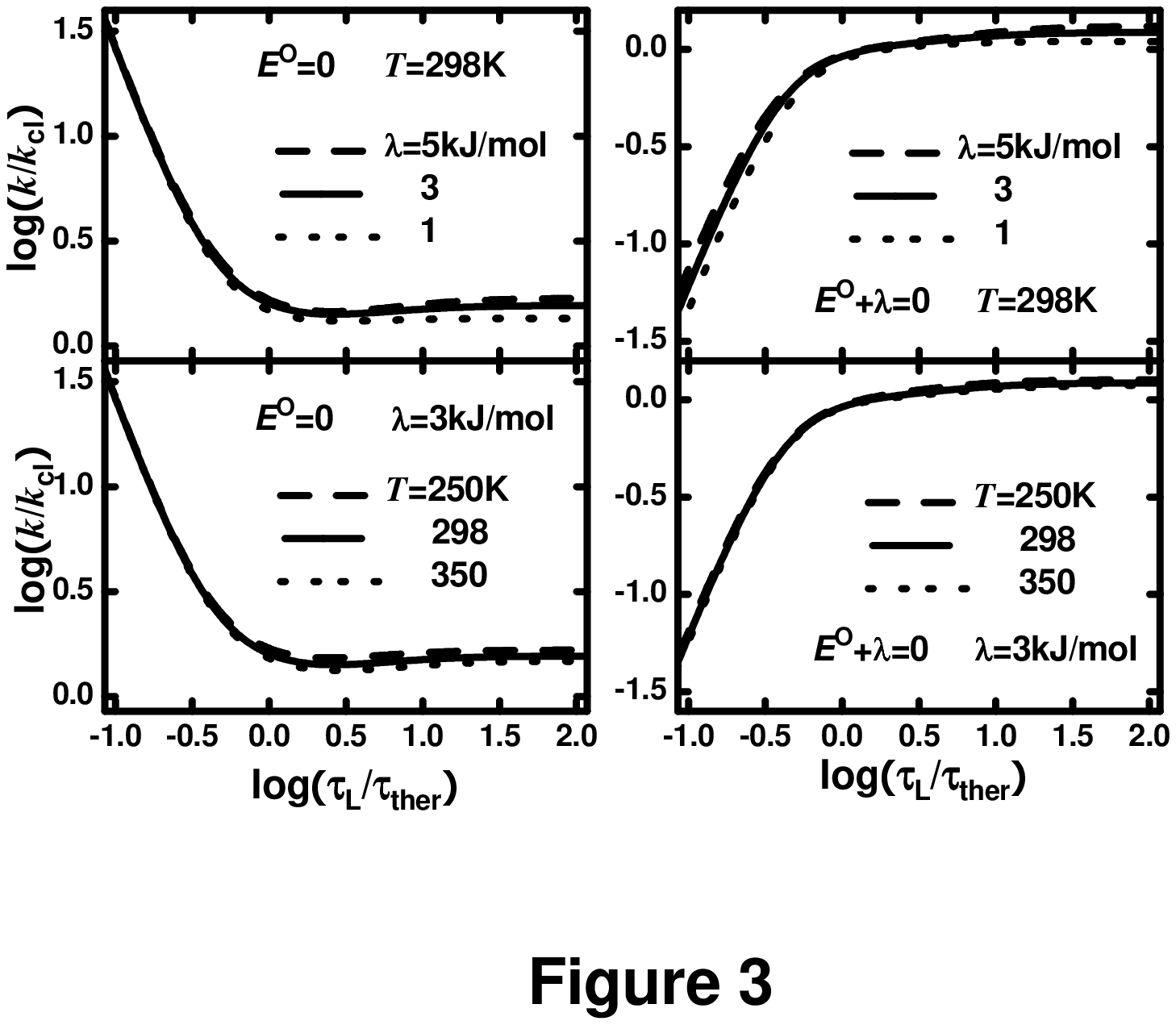}}
\centerline{\includegraphics[width= 0.90\columnwidth,angle=0]{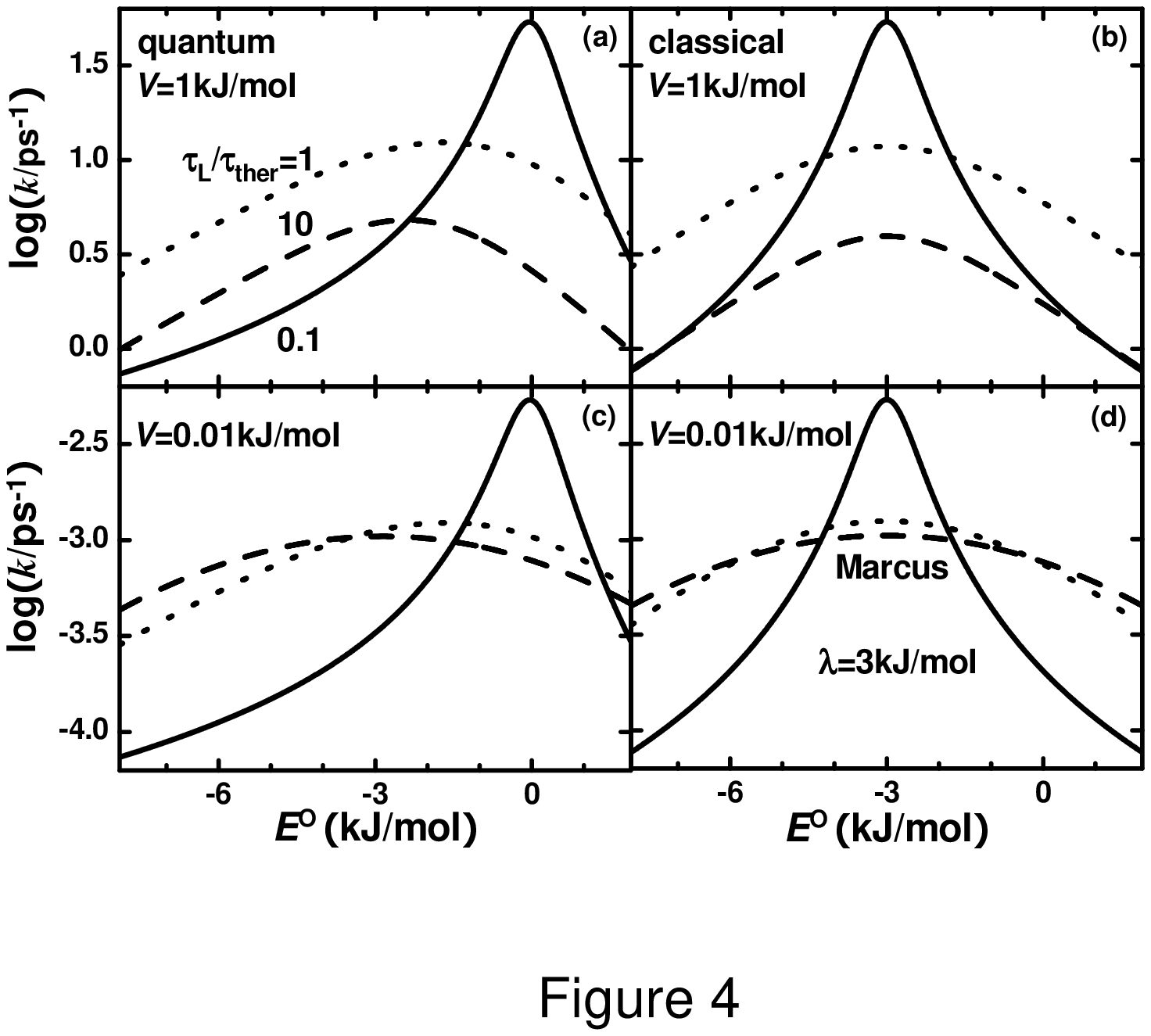}}
\centerline{\includegraphics[width= 0.90\columnwidth,angle=0]{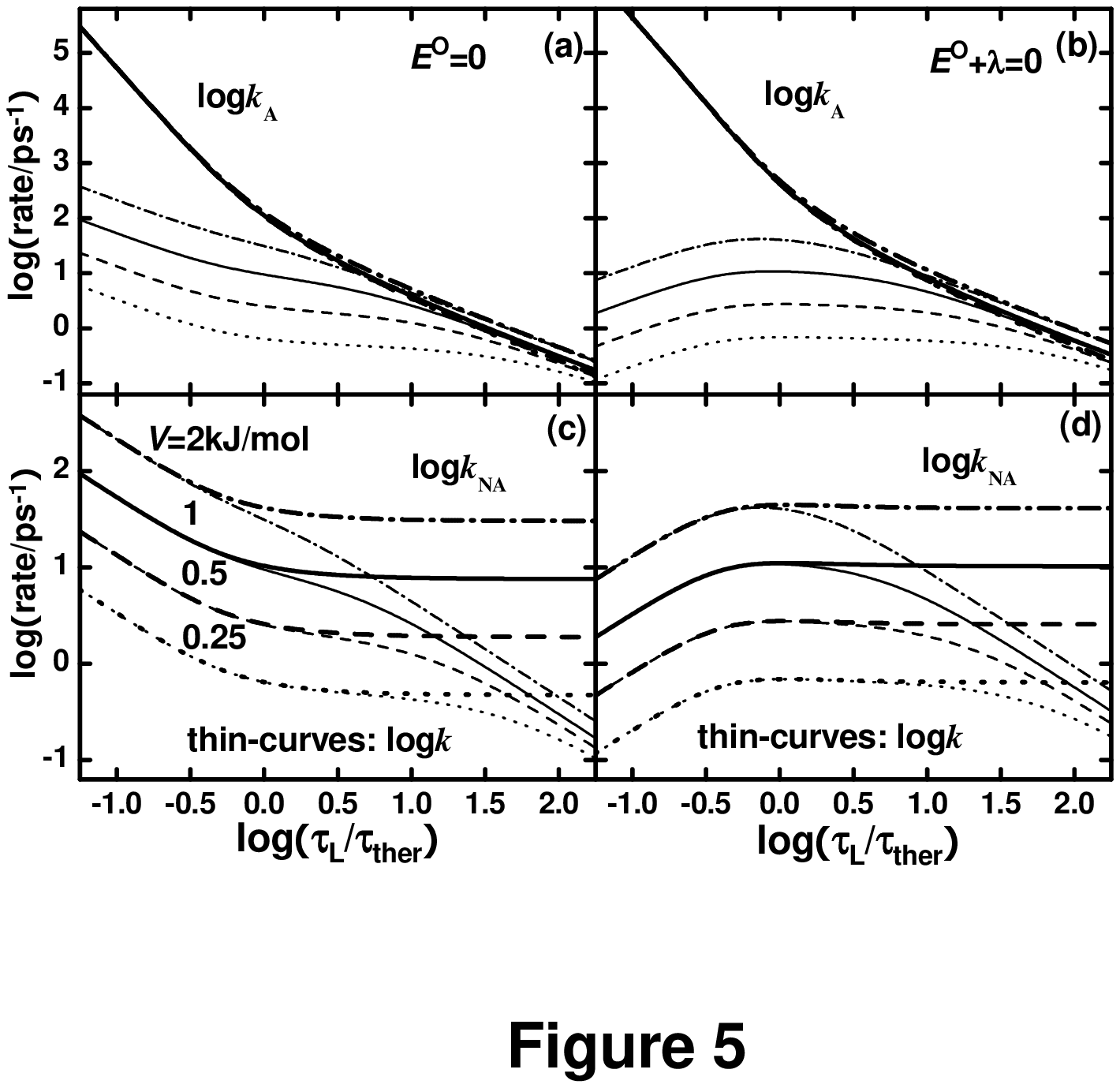}}
\centerline{\includegraphics[width= 0.90\columnwidth,angle=0]{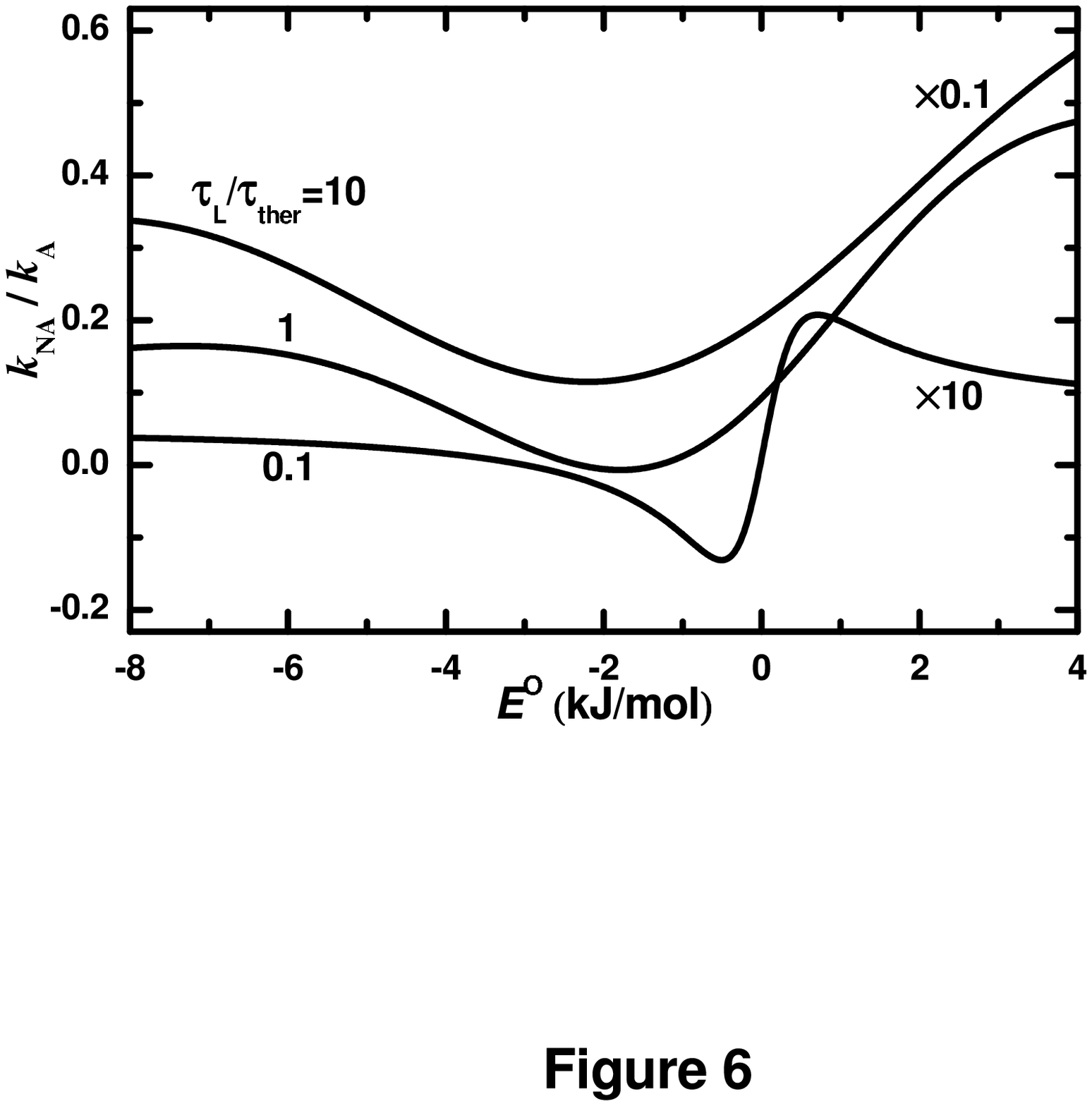}}
\centerline{\includegraphics[width= 0.75\columnwidth,angle=0]{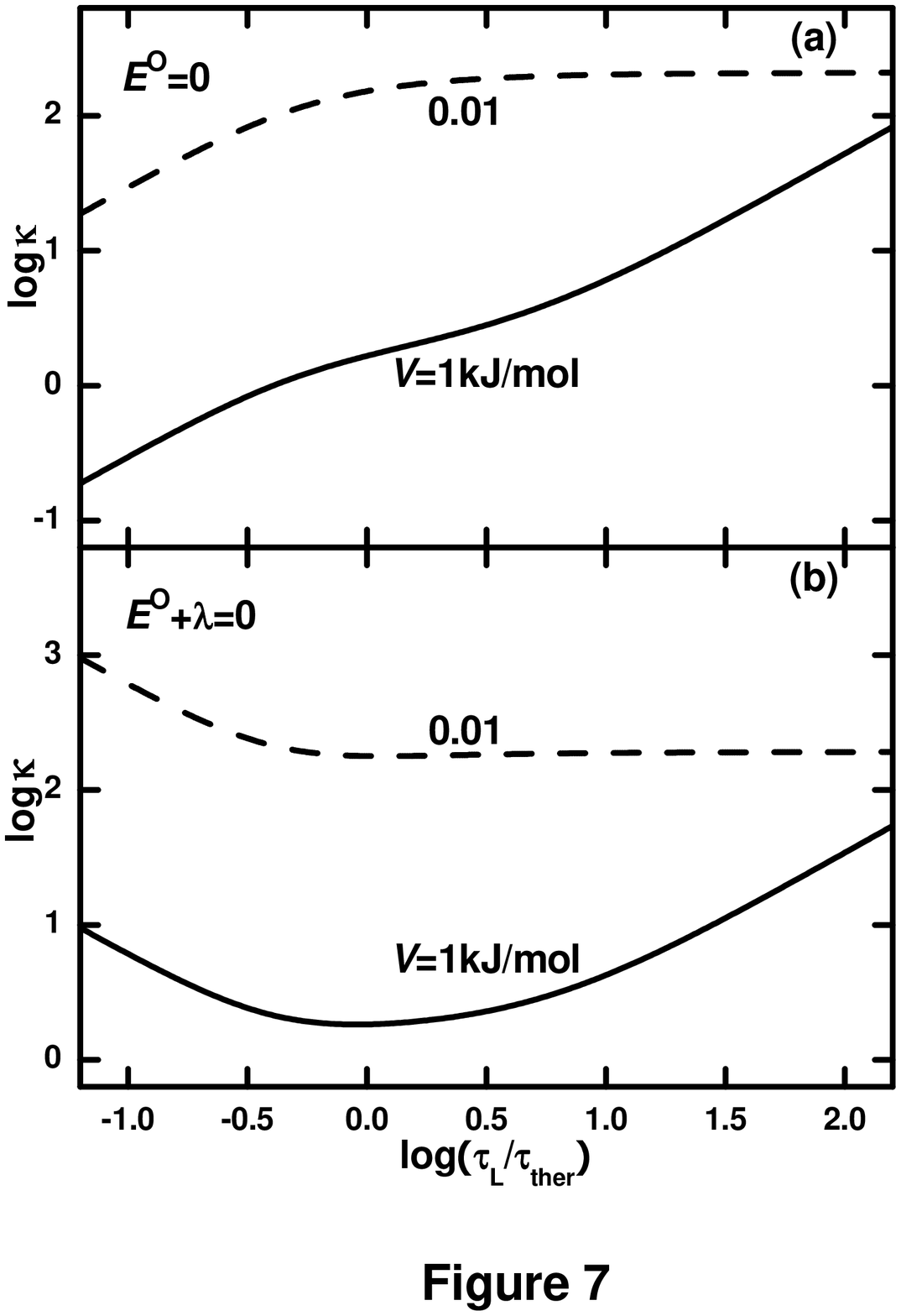}}
\centerline{\includegraphics[width= 0.90\columnwidth,angle=0]{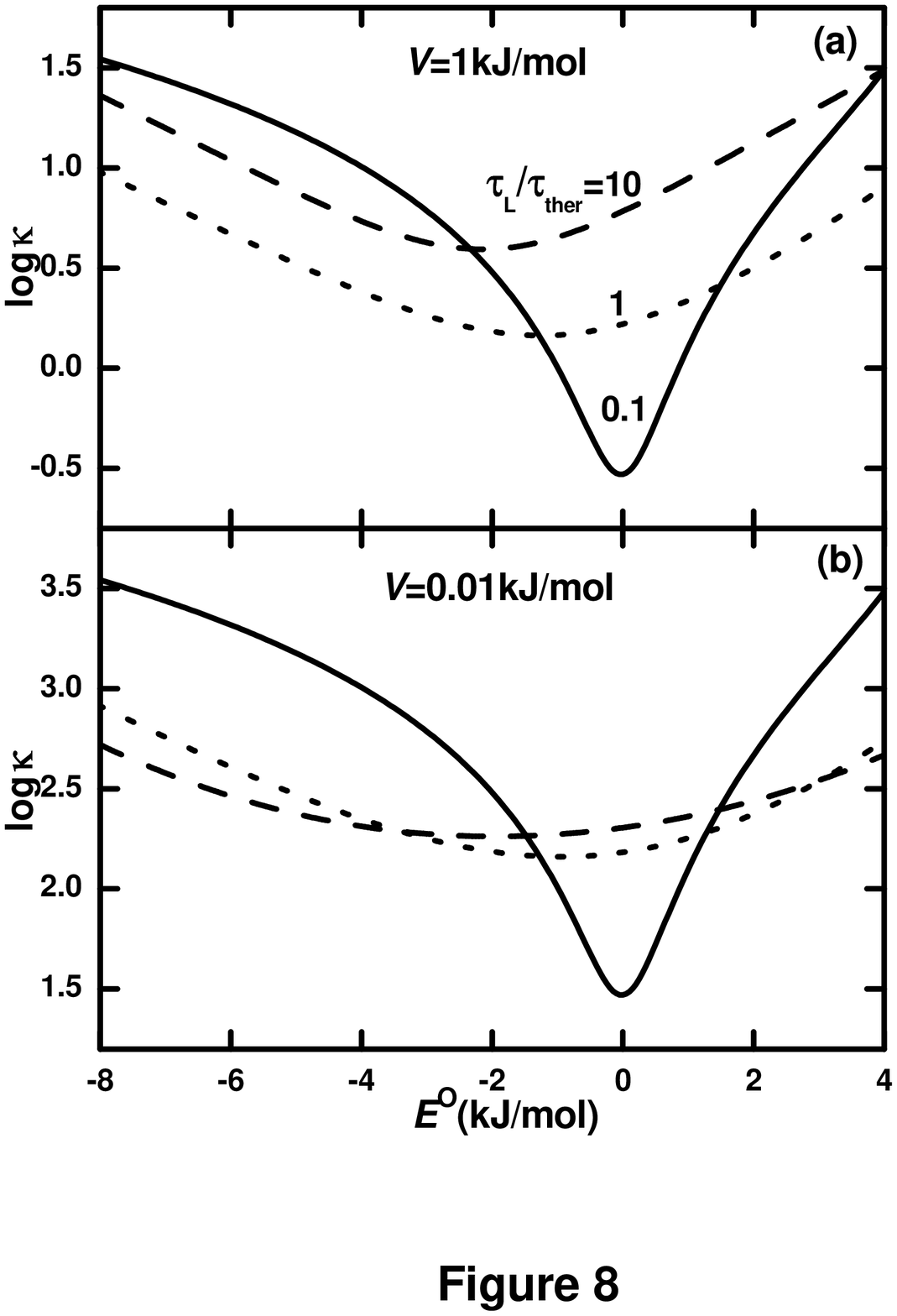}}
\centerline{\includegraphics[width= 0.90\columnwidth,angle=0]{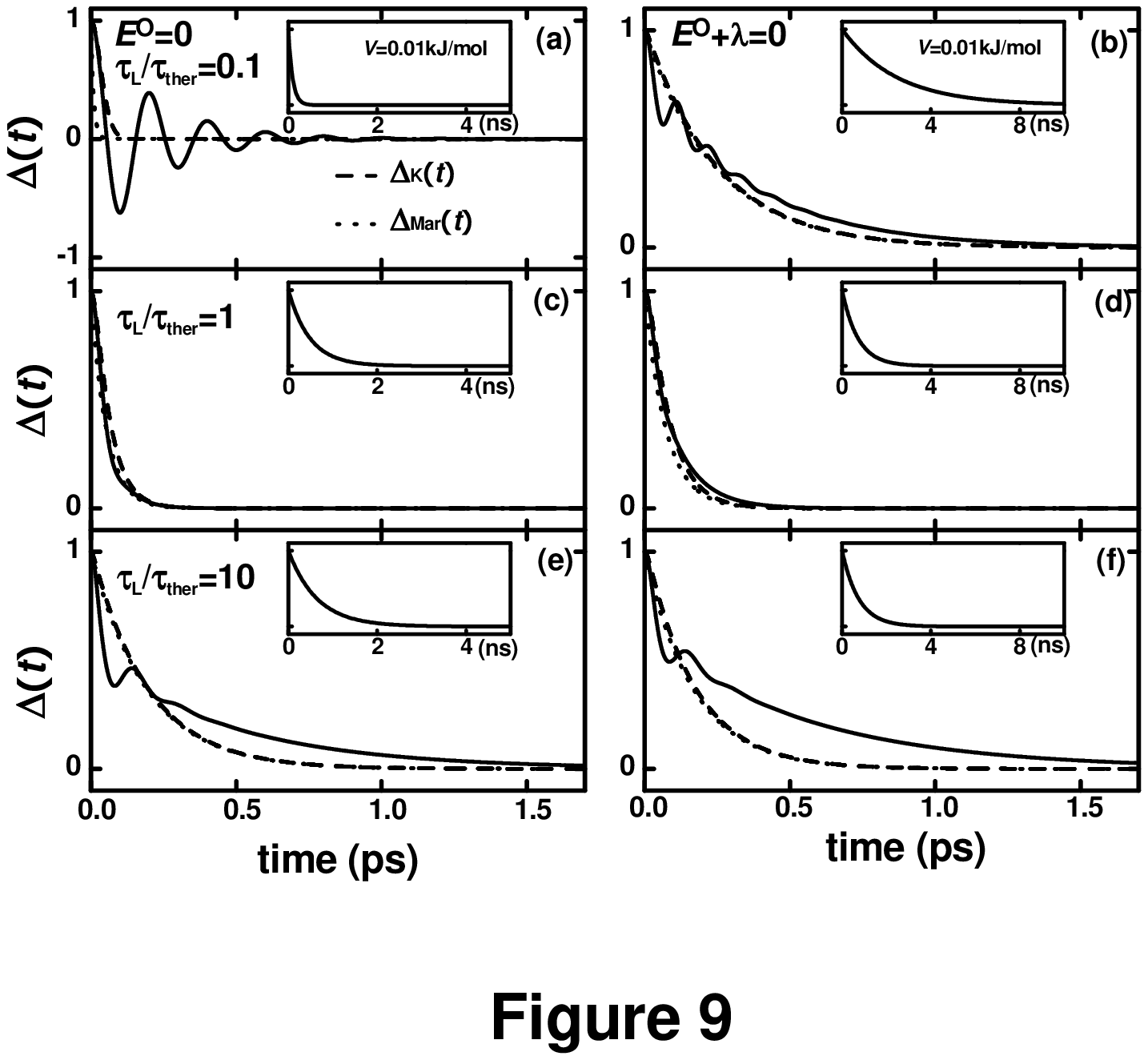}}
\centerline{\includegraphics[width= 0.90\columnwidth,angle=0]{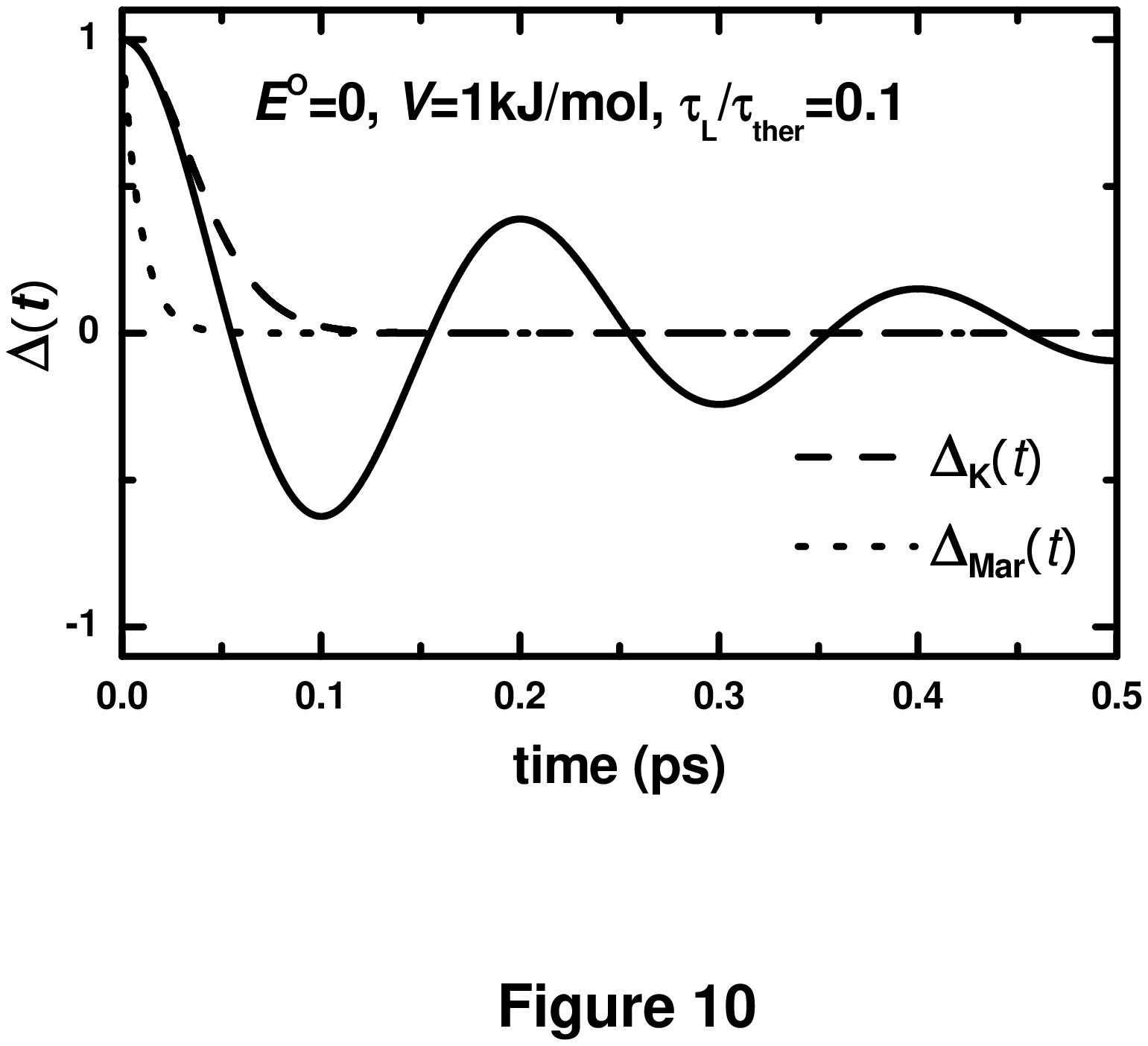}}
\end{center}

\end{document}